\documentclass[12pt]{article}
\usepackage{amsmath}
\usepackage{latexsym}
\usepackage{amsfonts}
\begin{document}

\vspace *{1in}

\begin{flushright}
UTEXAS-HEP-00-7
\end{flushright}

\vskip 1in

\begin{center}
{\Large \bf Brane Universe and Standard Cosmology}

\bigskip

Yungui Gong 

Physics Department, The
University of Texas at Austin, Austin, TX 78712
 
\bigskip
 
\vskip 0.5in
{\Large\bf Abstract}
\end{center}

Brane cosmology takes the unconventional form $H\sim \sqrt{\rho}$.
To recover the standard cosmology, we have to assume that
the matter density is much less than the brane tension. We
show that the assumption can be justified even near
the end of inflation if we fine-tune the coupling constant
of the inflaton potential. As a consequence, the standard
cosmology is recovered after inflation.

\bigskip


\pagebreak

\parindent=4ex 

Recently, brane cosmology \cite{bdl}-\cite{maartens}
 has attracted much attention.
In this scenario, ordinary matter is confined in our four
dimensional world, while gravity can propagate in the other
extra dimensions. As in string theory, the $E_8\times E_8$
heterotic string theory is a limit of an eleven dimensional
M theory with the extra dimension to be an orbifold $S_1/Z_2$.
Randall and Sundrum found a static solution with flat
3-brane to the five dimensional Einstein equation by assuming
a negative bulk cosmological constant \cite{rs995}\cite{rs996}. 
In this setup, our
universe is a 3-brane residing at the boundary of the fifth
orbifold $S_1/Z_2$. In \cite{bdl}, the authors find that the brane universe
expands according to the unconventional law $H\sim \rho$. Later,
it was found that the expansion rate could take the conventional
form $H\sim \sqrt{\rho}$ provided that the matter density
is much less than the cosmological constant in the brane
\cite{cgkt}\cite{cgs}\cite{kkop}.
Due to the success of the standard cosmology, we would like
to have standard cosmology for most of the history of the
universe. We will show that it is the case if we fine-tune
the coupling constant of the inflaton potential.

We consider a five dimensional spacetime with an
orbifold fifth dimension. The two 3-branes are at
$y=0$ and $y=L$. The Born-Infeld action for a $p$-brane is
$$S=-T_p\int d^{p+1}x\sqrt{{\rm det}(G_{\mu\nu}+B_{\mu\nu}
+2\pi\alpha' F_{\mu\nu})}.$$
If $B_{\mu\nu}=F_{\mu\nu}=0$, then the brane action
becomes $S=-T_p\int d^{p+1} x\sqrt{-g}$ which is just
the action of a cosmological constant. 
The five dimensional Einstein action with two brane boundaries
is
$$S=\int_{-L}^L dy\int d^4 x\sqrt{-g}\left({R\over 2\kappa^2_5}
-\Lambda\right)+\int d^4 x\sqrt{-g_1}(-T+{\mathcal L}_m)
+\int d^4 x\sqrt{-g_2}(-T_*+{\mathcal L}_{*m}).$$

For an isotropic and homogeneous flat brane
universe embedded in five dimensions with the
metric 
$$ds^2=-N^2(t,y)dt^2+a^2(t,y)\delta_{ij}dx^i dx^j+B^2(t,y)dy^2,$$
the components of the five dimensional Einstein tensor are
\begin{gather}
G_{00}={N^2\over B^2}\left[-3{a''\over a}+3{a'\over a} 
{B'\over B}-3\left({a'\over a}\right)^2\right]+3\left(
{\dot{a}\over a}\right)^2+3{\dot{a}\over a}{\dot{B}\over B},\\
G_{55}=3{a'\over a}\left({a'\over a}+{N'\over N}\right)
+3{B^2\over N^2}\left[{\dot{a}\over a}{\dot{N}\over N}-
\left({\dot{a}\over a}\right)^2-{\ddot{a}\over a}\right],\\
G_{05}=3{a'\over a}{\dot{B}\over B}+3{N'\over N}{\dot{a}\over a}
-3{\dot{a}'\over a},\\
\begin{split}
G_{ii}=&{a^2\over B^2}\left[{N''\over N}+2{a''\over a}
+\left({a'\over a}\right)^2+2{a'\over a}{N'\over N}
-2{a'\over a}{B'\over B}-{N'\over N}{B'\over B}\right]\\
&+{a^2\over N^2}\left[-{\ddot{B}\over B}-2{\ddot{a}\over a}
-\left({\dot{a}\over a}\right)^2+2{\dot{N}\over N}
{\dot{a}\over a}-2{\dot{B}\over B}{\dot{a}\over a}+
{\dot{N}\over N}{\dot{B}\over B}\right],
\end{split}
\end{gather}
where ${\dot a} \equiv da/dt$, ${\ddot a}\equiv d^2 a/dt^2$
and $a'\equiv da/dy$.
The components of the energy momentum tensor are
\begin{gather*}
T_{00}=N^2\Lambda +{\delta(y)\over B(t,0)}N^2(t,0)(T+\rho)
+{\delta(y-L)\over B(t,L)}N^2(t,L)(T_*+\rho_*),\\
T_{ii}=-a^2\Lambda +{\delta(y)\over B(t,0)}a^2(t,0)(-T+p)
+{\delta(y-L)\over B(t,L)}a^2(t,L)(-T_*+p_*),\\
T_{55}=-\Lambda B^2,\qquad T_{05}=0.
\end{gather*}
Here we assume that the matter on the branes is
the perfect fluid
$T_{\mu\nu}=(\rho+p)U_\mu U_\nu + p g_{\mu\nu}$.
Let $a_0=a(t,0)$, $N_0=N(t,0)$ and $B_0=B(t,0)$. 
The delta functions in $T_{00}$ and $T_{ii}$ give us
the jump conditions on $a'$ and $N'$ because of the
Einstein equation $G_{MN}=\kappa^2_5 T_{MN}$. The jump
conditions are
\begin{gather}
\label{jump1}
{3[a']\over a_0 B_0}=-\kappa^2_5 (T+\rho),\\
\label{jump2}
{3[N']\over N_0 B_0}=\kappa^2_5 (2\rho+3p-T).
\end{gather}
We use the notation $[f]\equiv f(0_+)-f(0_-)$ and $\{f\}\equiv
(f(0_+)+f(0_-))/2$. The $Z_2$ symmetry gives us
${a'}={N'}={B'}=0$.
The average of $G_{55}$ at the brane boundary gives the
equation of brane scale factor as follows:
\begin{equation}
\label{univ1}
-{\dot{a}_0\over a_0}{\dot{N}_0\over N_0}
+{\ddot{a}\over a_0}+\left({\dot{a}_0\over a_0}\right)^2
=N_0^2\left[{\kappa^4_5\over 18}T^2+{\kappa^2_5\over 3}\Lambda
+{\kappa^4_5\over 36}T(\rho-3p)-{\kappa^4_5\over 36}\rho
(\rho+3p)\right].
\end{equation}
The junction of $G_{05}$ at the brane boundary gives the
energy conservation in the brane
\begin{equation}
\label{univ2}
\dot{\rho}+3{\dot{a}_0\over a_0}(\rho+p)=0.
\end{equation}
The average and junction of other components give the
equations for $[B']$, $\dot{B}_0$ and $\dot{N}_0$.
Equations (\ref{univ1}) and (\ref{univ2}) are all what
we need to study the brane cosmology. In other
words, these boundary conditions give us enough
information about our world and we do not need
to solve the five dimensional bulk Einstein equation.
Define
$d\tau=N_0 dt$ and $H(\tau)=da_0/d\tau$. In terms
of the proper time $\tau$,
equations (\ref{univ1}) and (\ref{univ2}) become
\begin{gather}
\label{univ3}
{d^2 a_0/d\tau^2\over a_0}+H^2={\kappa^4_5\over 18}T^2
+{\kappa^2_5\over 3}\Lambda
+{\kappa^4_5\over 36}T(\rho-3p)-{\kappa^4_5\over 36}\rho
(\rho+3p),\\
\label{univ4}
{d\rho\over d\tau}+3H(\rho+p)=0.
\end{gather}
From now on, we change notation $\dot{f}=df/d\tau$.
Equation (\ref{univ3}) can be rewritten as the familiar
form \cite{kim}\cite{kraus}\cite{ftw}
\begin{gather}
\label{braneuniv}
H^2={\kappa^4_5\over 18}T\rho+{\kappa^4_5 \over 36}\rho^2
+{\kappa^2_5\over 36}T^2 + {\kappa^2_5 \over 6}\Lambda,\\
\label{braneuniv1}
{\ddot{a}_0\over a_0}=-{\kappa^4_5\over 36}T(\rho+3p)
-{\kappa^4_5\over 36}\rho(2\rho+3p)+{\kappa^4_5\over 36}T^2
+{\kappa^2_5\over 6}\Lambda.
\end{gather}
In general, we can get
\begin{gather*}
{\ddot{a}\over a}+\left({\dot{a}\over a}\right)^2=-\alpha\rho
(\rho+3p)+{\beta\over 2}(\rho-3p)+2\Lambda,\\
{\ddot{a}\over a}=-\alpha\rho(2\rho+3p)-{\beta\over 2}(\rho+3p)
+\Lambda,
\end{gather*}
from the following equations
\begin{gather*}
H^2=\alpha\rho^2+\beta\rho +\Lambda,\\
\dot{\rho}+3H(\rho+p)=0.
\end{gather*}
From equation (\ref{braneuniv}), we see that the standard
cosmology is recovered if $\rho\ll 2T$ and $\Lambda\sim
-\kappa^2_5 T^2/6$. The reason why we do not want the exact
cancellation between the bulk cosmological constant and the
brane tension is that these two terms may give the small effective
cosmological constant in four dimensions to be consistent
with the current observation. 
During the early times, their contributions are negligible.
The effective Newton's
constant in four dimensions is $G=\kappa^4_5 T/48\pi\sim
T/M^6$, where $M$ is the five dimensional Planck mass.
In this letter, we would like to show that
that the standard cosmology is recovered near
the end of inflation. Now we focus
on the inflationary phase. From 
equation (\ref{braneuniv1}), we know that inflation occurs as long
as $2\rho+3p<0$. Suppose inflation is driven by a single scalar 
field $\phi$, $\rho=\dot{\phi}^2/2+V(\phi)$ and
$p=\dot{\phi}^2/2-V(\phi)$. Inflation ends when $\dot{\phi}_e^2\sim
V(\phi_e)$. 

By the slow-roll approximation,
\begin{equation}
\label{approx}
\dot{\phi}^2\ll V(\phi),\qquad
|\ddot{\phi}|\ll |3\,H\dot{\phi}|,
\end{equation}
the equations (\ref{univ4}) and (\ref{braneuniv}) become
\begin{gather}
\label{bruniv1}
H^2\approx {\kappa^4_5  \over 18}T V(\phi)+{\kappa^4_5\over 36}V^2(\phi),\\
\label{bruniv2}
3H\dot{\phi}\approx -V'(\phi).
\end{gather}
The consistency conditions for the above 
approximations (\ref{approx}) are
\begin{gather}
\label{consist1}
{\bar \epsilon} \equiv {4\over \kappa^4_5}{V^{\prime 2}(\phi)\over
2T V^2(\phi)+ V^3(\phi)} \ll 1,\\
\label{consist2}
{\bar \eta} \equiv {4\over \kappa^4_5} \biggl|{V''(\phi)\over 
V^2(\phi)+2TV(\phi)}
-{V(\phi)V^{\prime 2}(\phi)+T V^{\prime 2}(\phi)\over [
V^2(\phi)+2TV(\phi)]^2}\biggr|\ll 1.
\end{gather}
To be more specific, we first choose the chaotic potential
$V(\phi)=\lambda_p\phi^p/p$.
From equations (\ref{consist1}) and (\ref{consist2}),
inflation ends when $\phi_e\sim 2p/(\kappa^2_5\sqrt{T})$.
So 
$$V(\phi_e)\sim {\lambda_p\over \kappa^{2p}_5 T^{p/2}}.$$
If 
$$\lambda_p < 2 \kappa^{2p}_5 T^{1+p/2}\sim 
{M^6\over m^{p+2}_{\rm pl}},$$
then $V(\phi_e)< 2T$. For example, let $p=4$.
The density fluctuation during inflation from the standard
cosmology gives $\lambda_4\sim 10^{-14}$. Here we require
$\lambda_4 \sim (M/m_{\rm pl})^6$. This condition
is not difficult to be satisfied. 

As another example, we take the exponential potential
$V(\phi)=V_0e^{-\phi/\phi_0}$. For this potential,
inflation ends when 
$$V(\phi_e)\sim {4\over \phi_0^2\kappa^4_5} -2T.$$
If
$${m_{\rm pl}\over \sqrt{48\pi}}<\phi_0<
{m_{\rm pl}\over \sqrt{24\pi}},$$
then we have $V(\phi_e)<2T$. 

During radiation dominated era, $\rho\propto a^{-4}$. Soon
after inflation $\rho$
will be much less than $T$. Therefore, the standard
cosmology is recovered during the radiation and matter dominated
epochs after inflation.


\begin{thebibliography}{d1}
\bibitem{bdl} P. Bin\'{e}truy, C. Deffayet and D. Langlois, 
hep-th/9905012, Nucl. Phys. B{\bf 565} (2000) 269.
\bibitem{rs995} L. Randall and R. Sundrum, hep-ph/9905221, 
Phys. Rev. Lett. {\bf 83}, (1999) 3370.
\bibitem{nihei} T. Nihei,  hep-ph/9905487, Phys. Lett. B{\bf 465} (1999) 81.
\bibitem{rs996} L. Randall and R. Sundrum, hep-ph/9906064, 
Phys. Rev. Lett. {\bf 83}, (1999) 4690.
\bibitem{cgkt} C. Cs\'{a}ki, M. Graesser, C. Kolda and J. Terning, 
hep-ph/9906513, Phys. Lett. B{\bf 462} (1999) 34.
\bibitem{cgs} J. M. Cline, C. Grojean and G. Servant, hep-ph/9906523,
Phys. Rev. Lett. {\bf 83} (1999) 4245.
\bibitem{kim} H.B. Kim and H.D. Kim, hep-th/9909053, 
Phys. Rev. D {\bf 61} (2000) 064003.
\bibitem{kkop} P. Kanti, I.I. Kogan, K.A. Olive and M. Pospelov, 
hep-ph/9909481, Phys. Lett. B{\bf 468} (1999) 31.
\bibitem{kraus} P. Kraus, hep-th/9910149, J. High Energy Phys. 
{\bf 9912} (1999) 011.
\bibitem{ftw} E.E. Flanagan, S.H. Tye and I. Wasserman, hep-ph/9910498.
\bibitem{vollick} D.N. Vollick, hep-th/9911181.
\bibitem{stw} H. Stoica, S.H. Tye and I. Wasserman, hep-th/0004126.
\bibitem{maartens} R. Maartens, hep-th/0004166.

\end{thebibliography}
\end{document}